# The Associative Multifractal Process: A Novel Model for Computer Network Traffic Flows

G. Millán, *Member, IEEE*, G. Lefranc, *Senior Member, IEEE*, and R. Osorio-Comparán

*Abstract*—A novel constructive mathematical model based on the multifractal formalism in order to accurately characterizing the localized fluctuations present in the course of traffic flows today high-speed computer networks is presented. The proposed model has the target to analyze self-similar second-order time series representative of traffic flows in terms of their roughness and impulsivity.

## I. INTRODUCTION

IN high-speed computer network traffic modeling, the robust characteristics present in most network structures, such as active users, applications, protocols, and topology [1]–[4], are considered. To obtain a parsimonious model of the traffic, it is avoided to describe localized fluctuations [5], [6], that are very sensitive to the conditions of the network [7]. It is argued that, in the extrapolation of the traffic observation scales [4], it can be carried out because, in the analysis of the time series of real life, they show cut-off points of lower and upper scale [1], [7], sometimes separated by several orders of magnitude [7]. This point of view adopts, as an axiom, that it is not worth thinking about dichotomous behaviors [8] and even less about theories focused on the alteration of transversal phenomena as a result of their exacerbated structural reductivism.

Traffic modeling, based on conventional telephone systems, is based on two assumptions: independence between arrival times of successive frames and exponential durations in the use of resources [1], [8].

These assumptions imply restricting stochastic processes to a Poisson universe. These considerations have been useful to designers and analysts in planning capabilities and predicting performance in systems. However, there are many cases in the real world [7] in which it has been observed that the predicted results, from a queue analysis [9], differ significantly from the performance observed in reality, and that this discrepancy has its origin in that the processes of traffic flows frequently shows far-reaching variations at different time-scales [1], [7], [8].

In the Poisson models [10], which have no memory or have short-range memory, they expose the flows of traffic bursts on much smaller timescales [7], [8]. As a result, overly optimistic forecast about performance are obtained due to the use of distributions with finite variance to characterize the periods of presence and absence of packet bursts [7].

Two modelling streams coexist; a conventional one, based on Poisson processes, and a self-similar one that accepts long-range dependency (LRD) as an inherent characteristic of data traffic flows in current high-speed computer networks [4], [6].

Self-similarity and LRD are not synonymous and formally one condition does not necessarily imply the other, but from the point of view of traffic modelling, self-similar processes are characterized by their invariance to scale changes and their ability to exhibit long-range correlations, that is, LRD [11].

The advantages of the LRD approach are evident over the conventional one [8], especially in reference to the more than questionable assumption of independence both between successive shipments and arrivals in traffic flows in computer networks. However, there is a certain number of researches that show a lack of consensus about the scope of applicability of self-similar models and the impact that the LRD has on the performance of communication systems [1], even though their number is quite small: Their conclusions should be carefully analyzed since they reveal a fundamental question for the multifractal analysis, which can be stated as: since traditional queuing models are not capable of evidencing self-similarity [7], their validity to predict yields would be supported by demonstrating that the self-similarity does not have a truly measurable impact on them. And even more, if it is shown that self-similar models fail to consider the impact of important individual parameters in each particular case of network or communications system [1], [7].

There is another even more critical question that points directly to the center of self-similar modelling: the validity of the Hurst exponent ($H$) as a descriptor of the traffic LRD. In [12] it is shown that, through an exhaustive queue analysis applied to a representative time series of Ethernet traffic traces $H$ does not provide an accurate prediction of queue performance for a given LRD traffic [12]. And furthermore, its behavior is monotonic with respect to the presence or absence of packet bursts, if the original series is broken down into other smaller ones, which implies that $H$ also does not serve to characterize the importance of the groupings within the total traffic, thus dismissing aggregation as a method of analysis [1], [12].

G. Millán is with the Universidad San Sebastián, Puerto Montt 5501842 Chile (e-mail: ginno.millan@usach.cl).
G. Lefranc is with the Pontificia Universidad Católica de Valparaíso, Valparaíso 2362804 Chile (e-mail: gaston.lefranc@pucv.cl).
R. Osorio-Comparán is with the IIMAS, Universidad Nacional Autónoma de México (e-mail: roman@unam.mx).

These questions find an answer in the analysis of behavior on a small scale [7], [8], a product that, quite the opposite of what happens with the large-scale analysis, this considers the localized fluctuations in the course of the samples of the time series representative of the traffic flows of traffic and does not discriminate them in favor of more robust singularities [7].

Specifically, it is about the use of the multifractal formalism understood as a set of coordinated fractal varieties capable of characterizing the evolution of a phenomenon that exhibits self-similarity at different observation scales to study traffic flows [8].

## II. Structure of the Proposed Model

The original On/Off model provides detailed information on the generation of traffic from both individual and aggregated sources [13].

Then, it is proposed a generalization of its properties through the introduction of the multifractal On/Off model concept.

### A. The On/Off Model

An On/Off process alternates between two states: On during which a source generates traffic with an $A_j$ rate and Off during which the source does not generate traffic.

Then, being $X_j$ and $Y_j$ the durations of the $j$th On and Off states, respectively, an conventional On/Off process is defined by the expression

$$S(t) = \sum_{j=0}^{\infty} A_j 1_{[S_j, S_j + X_j)}(t), \quad \forall t \geq 0, \ j \in \mathbb{Z}, \quad (1)$$

with $I_j(t) = 1_{[S_j, S_j + X_j)}(t)$, then

$$S(t) = \sum_{j=0}^{\infty} A_j I_j(t), \quad \forall t \geq 0, \ j \in \mathbb{Z}, \quad (2)$$

where is verified that

- $S_j$: Regenerative point [14]. Indicates the occurrence of the $j$th state On, defined by

$$S_j = S_0 + \sum_{i=1}^{j}(X_i + Y_i), \quad \forall j \geq 1, \quad (3)$$

where $S_0$ is the beginning of the observation time and it is accepted that $S_0 = 0$.

- $I_j(t)$: Indicator function; $I_j(t) = 1$ for $t \in [l_1, l_2)$, where $l_1$ and $l_2$ are fixed values according to the On states.

In the On/Off model, each $X_j$, $Y_j$ are assumed independently and identically distributed according to whether the hyperbolic tail distribution presents or not finite variance.

The Hurst exponent of an On/Off model according to (1) is given by the expression [15]

$$H = \frac{3 - \min\{\alpha_0, \alpha_1\}}{2}, \quad (4)$$

where $\alpha_0$ and $\alpha_1$ are the tail indices of the durations of the On and Off states, respectively. If the durations of the On and/or Off states have finite variance, the index of their queue is taken with a value of 2 when (2) is applied.

The original On/Off process given by (1) is self-similar if $1/2 < H < 1$, which implies that the shorter duration of one of the states (On or Off) presents a hyperbolic tail distribution so that the On/Off is self-similar.

### B. The On/Off Multifractal Model

Let $\Theta(t)$ be a conventional On/Off process given by (1). It is said that $\Theta(t)$ is an Associative Multifractal Process, $\Theta_{AM}(t)$, if the durations of the On and Off, $X_j$ and $Y_j$ states, respectively, are hyperbolically distributed with density functions given by $f_1(x), f_0(x)$, respectively, and such that

$$\Theta_{AM}(x): f_i(x) = x^{-(\alpha_i + 1)}, \quad \alpha_i \in (1,2), \ i = 0,1, \quad (5)$$

where $f_0(t) = f_1(t) = 0$ for $t = 0$, and time means of On and Off states, both finite, are given by $\mu_1 = \mathrm{E}\{X_j\}$ and $\mu_0 = \mathrm{E}\{Y_j\}$.

*1) Comments About (5)*

In the statement of (5) it is necessary to emphasize that

1. The expected value of a stochastic process governed by (5) is given by $\mathrm{E}\{\Theta_{AM}(x)\} = \mu_1/(\mu_0 + \mu_1)$.
2. It is assumed that the traffic generation rate is constant in the individual On states, for the different compound On states, that is, $A_j = c, \forall j$.
3. Applying the previous condition in (2) we obtain

$$\Theta_{AM}(t) = c\sum_{j=0}^{\infty} I_j(t), \quad \forall t \geq 0. \quad (6)$$

4. According to [16] the Power Spectral Density (PSD) for (1) is given by

$$\mathrm{PSD}[S(t)] = \mathrm{E}\{S(x)\}\delta\left(\frac{\omega}{2\pi}\right) + \frac{2/\omega^2}{\mu_0 + \mu_1}\mathrm{Re}\left\{\frac{[1 - F_0(-j\omega)][1 - F_1(-j\omega)]}{1 - F_0(-j\omega)F_1(-j\omega)}\right\}, \quad (7)$$

where $F_0(-j\omega)$ and $F_1(-j\omega)$ are the characteristic functions of $f_0(x)$ and $f_1(x)$, respectively, and $\delta(\cdot)$ the Dirac function. Thus, for (5), (7), replacing $\mathrm{E}\{\Theta_{AM}(x)\}$ according to the first consideration, becomes

$$\mathrm{PSD}[\Theta_{AM}(t)] = \frac{\mu_1}{\mu_0 + \mu_1}\delta\left(\frac{\omega}{2\pi}\right) + \frac{2\omega^{-2}}{\mu_0 + \mu_1}\mathrm{Re}\left\{\frac{[1 - F_0(-j\omega)][1 - F_1(-j\omega)]}{1 - F_0(-j\omega)F_1(-j\omega)}\right\}. \quad (8)$$

Result that highlights the coherence between the structure of the $\Theta_{AM}(t)$ processes and a well-known theory such as the multifractal theory [17], in addition to accounting for the flexibility of the original On/Off approach of [13] as regards aggregate traffic representation is concerned.

5. An important subclass distribution that shows a regular variation are the hyperbolic-tailed distributions, which have a survival function according to

$$F_s(x) = Cx^{-\beta}G(x), \quad \text{when } x \to \infty, \quad (9)$$

where $G(x)$ is a slowly varying function and $0 < \beta < 2$.

With the above for (5) it is verified that: If the durations of the On and Off periods are Zipf distributed with reliability functions given by $F_{S1}(x; \alpha_1, k_1)$, $F_{S0}(x; \alpha_0, k_0)$, respectively. Then, for $\omega \to 0^+$, from (8) we have the expression [18]

$$S(\omega) \approx W_0 \omega^{\alpha_0 - 2} + W_1 \omega^{\alpha_1 - 2}, \quad (10)$$

where $W_0, W_1 \in \mathbb{R}^+ - \{0\}$.

6. It is possible to simplify (10) without loss of generality as follows

$$S(\omega) \approx W\omega^{\alpha - 2}, \quad \text{where } \alpha = \min\{\alpha_0, \alpha_1\}. \quad (11)$$

7. From [19] it is known that a random process $X$ with finite second-order statistics is stationary with LRD in the sense of its autocorrelation, if its autocorrelation function, $R_{xx}(k)$, satisfies asymptotically that

$$\lim_{k \to \infty}\left[ R_{xx}(k)/k^{-\beta} \right] = C_A, \quad (12)$$

where $0 < \beta < 1$, with $\beta = 2 - 2H$; and $C_A > 0$.

An equivalent definition of LRD is described in [20] based on the process spectrum: a process has LRD if this spectrum satisfies

$$\lim_{\lambda \to 0}\left[ f(\lambda)/\left(C_S |\lambda|^{-\beta}\right) \right] = 1, \text{ with } C_S < 0. \quad (13)$$

8. Then, an On/Off process given by (5) Pareto distributed On and Off periods with tail indices $\alpha_1$ and $\alpha_0$, respectively, is LRD in the sense of (13) if

$$\lim_{\lambda \to 0}\left[ f(\lambda)/\left(C_S |\lambda|^{-\beta}\right) \right] = 1, \quad (14)$$

where $C_S \in \mathbb{R}^+ - \{0\}$ and $H$ given by (4).

III. THE ASSOCIATIVE MULTIFRACTAL PROCESS MODEL

In Section II the mathematical foundations that make up the structure of the model were delivered (see (5)). It corresponds to involve an $\Theta_{AM}(x)$ process with the sources that make up a high-speed computer network.

*A. General Modeling Condition*

Being $X_j$ and $Y_j$ the durations of the On and Off states, for modeling purposes the following conditions are assumed

1. The processes $X_j$ and $Y_j$ [21] are independent and identically distributed with reliability functions given by $F_{S1}(x; \alpha_1, k_1)$, $F_{S0}(x; \alpha_0, k_0)$, with $1 < \alpha_0 < 1$, $1 < \alpha_1 < 2$, respectively.
2. The traffic generation rate, $A_j$, during the processes $X_j$ is independent and identically distributed of Bounded Pareto with reliability function $F_{SB}(x; \alpha_B, k_B)$ independent of the durations of the On and Off states.

*B. Previous General Comments*

It is necessary to establish the following guidelines, which are valid in the general context of the model

1. The use of heavy-tail distributions for the durations of the On and Off states is based on the observations of [21], which suggest that the On and/or Off states can be very long with a high probability.
2. If the durations of the On and Off states are considered to be heavy-tail distributed with finite variance, the overlap of many sources will behave as short-range dependent traffic, which conflicts with the precepts and conclusions of [22], and therefore on with the basis of this research.
3. The Bounded Pareto distribution, in terms of its reliability function, is defined by [18]

$$F_{SB}(x;\alpha_B,k_B) = F_{SB}(x;\alpha_B,k_B)(1 - u_B(x - B))$$

$$= \begin{cases} (k_B/x)^{\alpha_B}, & k_B \leq x \leq B \\ 1, & x < k_B \\ 0, & x > B \end{cases} \quad (15)$$

where $u_B(\cdot)$ unit step function and $B$ cut-off limited imposed on the random variable.

Furthermore, from [18] the Bounded Pareto function has the following density probability function

$$\begin{aligned} f_{SB}(x;\alpha_B,k_B) &= f(x;\alpha_B,k_B)(1 - u_B(x - B)) + \\ &\quad F(x;\alpha_B,k_B)\delta(x - B) \\ &= f(x;\alpha_B,k_B)(1 - u_B(x - B)) + \\ &\quad k_B^{\alpha_B} x^{-\alpha_B}\delta(x - B), \end{aligned} \quad (16)$$

where $f_{SB}(\cdot)$, $F_{SB}(\cdot)$ Pareto probability density and reliability functions, respectively, and $\delta(\cdot)$ is the Dirac delta function.

4. The existence of cut-off limit $B$ causes the Bounded Pareto distribution, in contrast to the Pareto distribution, to have finite variance.

*C. Traffic Generated by a Source*

It is considered a source that alternates between On and Off states governed by (5). During the On states the traffic source generates traffic with constant rate $A_j$ and during the Off states the source it remains silent.

Then, the probability density function of the process $\Theta_{AM}(t)$ of (6) under the modeling conditions established in Subsection A, is defined by

$$\begin{aligned} f_{S\Theta_{AM}(x)}(x;\alpha,k_B,B,\mu_0,\mu_1) &= \mu_0(\mu_0 + \mu_1)^{-1}\delta(x) + \\ &\quad \mu_1(\mu_0 + \mu_1)^{-1} f_{SB}(x;\alpha_B,k_B), \end{aligned} \quad (17)$$

with $\mu_1$ and $\mu_0$ defined in (5).

*D. Traffic Generated by Multiple Sources*

*1) Recapitulation. General Modeling Considerations*

In the approach of (1) it is specified that $A_j$ is the rate with which the $j$th source generates traffic during On states, which have a duration determined by $X_j$ process.

This idea is later reflected in (5) originating (6) under the following three key modelling assumptions

1. The transmission rate in the individual On states is constant for the different compound On states, that is, $A_j = C \; \forall j$, to obtaining (6).
2. The durations of the On and Off states, given by $X_j$ and $Y_j$, respectively, are distributed with a heavy-tail with density functions $f_1(x)$ and $f_0(x)$, respectively, which gives rise to the associative multifractal process $\Theta_{AM}(x)$ from (5), where it is agreed that $f_0(t) = f_1(t) = 0$ for $t < 0$.
3. The means of the durations of the On and Off states; $\mu_1$ and $\mu_0$, ($\mu_1 = E\{X_j\}$, $\mu_0 = E\{Y_j\}$), respectively, are finite, a fact by which the expected value of a process governed by (5) is given by $E\{\Theta_{AM}(t)\} = \mu_1/(\mu_0 + \mu_1)$.

The previous assumptions in conjunction with the definition of a Bounded Pareto distribution in terms of its reliability function with upper cut-off limit $B$ (see (15)), give rise to (17), under the general modeling conditions given by Subsection A, expression that governs the probability density function of the $\Theta_{AM}(x)$ model when it is a source.

The model for a source in terms of its density function (see (17)) determines the regime that each source follows to inject traffic into the network. Then, for the case of multiple sources, it is necessary to consider the capacity of the link over which the composite traffic flows from $n$ sources are sent.

*2) Specific Modeling Assumptions*

The following guidelines are considered valid to develop the associative multifractal process model for *n* traffic sources

1. Let $M_t$ be the maximum allowable transmission rate for the link over which traffic flows from *n* sources are transmitted. That is, it is established that all individual transmission rates is limited by $M_t$ and this figure can never be exceeded, in other words, the link congestion situations are not feasible. Thus, each individual transmission is limited by a quantity $B_i$ such that $B_i \ll M_t$. The $B_i$ quantities are strictly different for the sources of the network since the transmission times of the sources are not the same and that some of them transmit simultaneously over more than one connection due to the operational nature of the TCP protocol.

2. Let $N$ independent identically distributed Bounded Pareto On/Off processes given by $\Theta_{AM(i)}(x)$, with $i = 1,\ldots, N$, each with a probability distribution function given by (17) and where it is verified $NB_i < M_t$. Then, the density function of $\Theta_{AM}(x)$ for *n* sources is given by the *N*-convolution of the individual density functions of the sources given by (17).

*3) Traffic Model for n Sources*

The *N*-convolution of (17) with $x > k_B$, $A_0 = \mu_0/(\mu_0 + \mu_1)$, and $A_1 = \mu_1/(\mu_0 + \mu_1)$ is approximated by the expression

$$f^N_{S\Theta_{AM}(x)}(x) \approx A_0^N \delta(x) + (1 - A_0^N)[\alpha(k_{B_{(N)}}/x)^{\alpha+1} \\ [u_B(x - k_{B_{(N)}}) - u_B(x - B)] + t_N(x)], \quad (18)$$

where

$$k_{B_{(N)}} = \exp\{(1 + \alpha_B)^{-1} \ln[(1 - A_0^N)^{-1}[(A_0 - A_0^N)k_{B_{(N-1)}}^{\alpha_B - 1} \\ A_0^{N-1} A_1 k^{\alpha_B + 1} + k_{B_{(N)}}(A_1 - A_1 A_0^{n-1})k_B^{\alpha_B + 1} k_{B_{(N-1)}}]]\},$$

and $t_N(x) = f^N_{S\Theta_{AM}(x)}(x)[u_B(x - B) - u_B(x - NB)]$.

IV. CONCLUSION

This letter, the mathematical development of a constructive model based on the multifractal formalism has been presented. The objective is to research the localized fluctuations, present in the representative time series of the traffic flows of current high-speed networks. This implies that the model focuses its attention on the study of the singularities of the samples that are located in subsets of minimum length within the time series.

The generation of traffic from a source is carried out under the guidelines of a conventional On/Off process but with the difference of the introduction of a maximum limit rate for its generation. The reason for introducing this limit rate is based mainly on the use of the TCP protocol reported by [23].

The traffic to multiple sources is treated as the convolution of traffic from individual traffic sources.

Finally, it can be said that the proposed model has bounded heavy-tailed and therefore when the number of traffic sources increases, the total traffic tends to be Gaussian, which reflects the consistency of the model with the behavior of real traffic.

It is noted that as the maximum rate limit for the generation increases, more active sources will be necessary for the traffic to become Gaussian behavior. This implies that in current high-speed networks the proposed model constitutes an upper limit of behavior.